\documentclass[pra,showkeys,preprint]{revtex4}

\usepackage{graphicx,booktabs,array}
\usepackage{amsmath}
\usepackage{amssymb}

\usepackage{graphicx,mathdots}
\usepackage{booktabs}
\usepackage{mathtools,tabularx}
\usepackage{makecell}
\setcellgapes{3pt}
\usepackage{graphicx,color}

\usepackage{graphicx,amssymb}
\usepackage{graphicx}

\begin{document}
\setcounter{page}{1}


\title{Scalar particle in the Kiselev-anti-de-Sitter black hole background}

\author{M. D. de Oliveira$^{1}$\footnote{Corresponding author. Email: dalpra.matheus@gmail.com} and Alexandre G. M. Schmidt$^{1}$}
\affiliation{Instituto de Ci\^encias Exatas, Universidade Federal Fluminense,\\ 
27213-145 Volta Redonda --- RJ, Brazil}


\begin{abstract}

In this work, we investigate the dynamics of a scalar particle in a black-hole spacetime surrounded by a cosmological constant and an anisotropic fluid. By solving the Klein--Gordon equation through a power series approach, we obtain quasibound states and the corresponding energy quasispectrum. We find that the quasispectrum can exhibit both real and imaginary components, depending on the model parameters, with the real component existing only under specific conditions. The numerical analysis shows that decreasing the cosmological constant and the anisotropic fluid parameter enhances the energy and increases the radiation in the region outside the event horizon, whereas stronger contributions from these surrounding fields tend to suppress the emitted radiation. We also analyze several asymptotic limits and recover the corresponding results for the Kiselev, de Sitter, and Schwarzschild black holes, providing consistency checks for our solutions. These results highlight the role of the surrounding matter distribution and cosmological contribution in modifying the dynamics and radiation of scalar fields in black hole spacetimes.

\end{abstract}

\maketitle

\section{Introduction}

The study of compact objects in strong gravitational fields, particularly black holes, plays a central role in contemporary theoretical physics, providing a framework in which general relativity, quantum field theory, and cosmology intersect. Exact solutions of Einstein's field equations, including the Schwarzschild and Kerr spacetimes \cite{kerr,gravitation}, have served as standard backgrounds for investigating a wide range of phenomena associated with strong gravitational fields. At the same time, observational results indicate that the present universe undergoes accelerated expansion, which is commonly attributed to a dark energy component \cite{riess,perlmutter,planck}. Despite its relevance to the large scale dynamics of the universe, the physical origin and fundamental nature of dark energy remain unresolved. This motivates the investigation of how dark energy inspired matter distributions can modify black hole geometries and their associated physical properties.

Among the theoretical frameworks proposed to account for dark energy, quintessence has received substantial attention. It is usually modeled through a dynamical scalar field with negative pressure \cite{ratra,caldwell,tsujikawa}, thereby providing a description that differs from the strictly constant energy density associated with the cosmological constant. In this context, Kiselev \cite{kiselev} derived an exact static and spherically symmetric solution of Einstein's equations for a black hole surrounded by a matter component described by a dark energy-like equation of state. The presence of this surrounding component introduces an additional contribution to the spacetime geometry and consequently modifies the gravitational structure of the black hole solution.

It is important, however, to clarify the interpretation of the matter source in the Kiselev solution. Although it is commonly described as a black hole surrounded by quintessence, Visser \cite{visser} demonstrated that the associated stress-energy tensor is intrinsically anisotropic and does not, in general, correspond to either a perfect fluid or conventional cosmological quintessence. The Kiselev configuration should therefore be understood more precisely as an effective anisotropic matter model exhibiting quintessence-like properties. Accordingly, we refer to it either as a Kiselev black hole or as a black hole surrounded by a quintessence-like anisotropic fluid. This distinction is relevant when relating the model to the phenomenology of dark energy and when interpreting the physical effects produced by the surrounding matter distribution.

Kiselev-type geometries have subsequently been employed in several investigations of black hole physics. The presence of the quintessence-like anisotropic component can modify thermodynamic quantities and the stability properties of the black hole system, including the Hawking temperature and entropy \cite{zhou,liu,ma}. Its influence also extends to observable features associated with photon propagation and perturbations of the gravitational background. For instance, modifications of black hole shadows and photon trajectories have been investigated in the presence of dark energy-like matter \cite{tsupko,amarilla,wei}. Quasinormal modes provide another important probe of the response of black holes to perturbations \cite{konoplya0,berti}, and studies of Kiselev configurations have shown that the surrounding matter can alter both the oscillation frequencies and damping characteristics of these modes. Furthermore, scalar-field dynamics in Kiselev backgrounds, including quasinormal modes, Hawking radiation, and the corresponding Hawking temperature, have been investigated in \cite{dalpraquint1,dalpraquint2}. These results indicate that the parameters describing the anisotropic matter distribution can have a significant impact on the geometry and on physical processes occurring in the vicinity of the black hole.

The introduction of a cosmological constant extends the Kiselev geometry to de Sitter and anti-de Sitter backgrounds, producing Kiselev--de Sitter and Kiselev--anti-de Sitter configurations. In these spacetimes, the combined effects of the anisotropic matter distribution and the cosmological background can modify the horizon structure as well as the behavior of fields propagating around the black hole \cite{gharderi}. In particular, Kiselev-type black holes embedded in AdS backgrounds have recently attracted attention in studies involving thermodynamic properties, geodesic dynamics, photon spheres, black hole shadows, and scalar-field quasinormal modes \cite{kiselevads2024,kiselevads2026}. Such developments provide motivation for a more detailed investigation of relativistic quantum fields in Kiselev--anti-de Sitter spacetimes.

The analysis of quantum fields in curved spacetime also provides a natural framework for investigating particle processes in strong gravitational fields. In particular, black holes in different gravitational settings have been shown to exhibit thermal emission as a consequence of quantum effects in the vicinity of the event horizon \cite{suzuki,hatsuda,bezerra1,vieira, vieira-2015, vieira-class-quantum-grav,hortacsu,Senjaya5,Senjaya6,Senjaya7,Senjaya8,Senjaya9,kokkotas}. This phenomenon, known as Hawking radiation, establishes an important connection between quantum theory and black hole thermodynamics \cite{hawking,bekenstein}. The characteristics of this radiation depend not only on the intrinsic parameters of the black hole but may also be affected by the surrounding spacetime geometry and matter content. Consequently, studying quantum fields in nontrivial cosmological environments can provide further insight into quantum processes occurring in strong gravitational regimes.

Motivated by these considerations, we investigate in this work the dynamics of a relativistic scalar particle in the spacetime of a Kiselev--anti-de Sitter black hole. We obtain the corresponding Klein--Gordon equation and analyze its radial sector, focusing on the structure of quasibound states and the associated energy quasispectrum. We further examine the scalar radiation in the exterior region of the event horizon and investigate how the parameters associated with the anisotropic matter distribution and the cosmological constant influence the resulting particle dynamics. In addition, we study several limiting configurations in order to establish connections with previously analyzed black hole geometries. These limits provide consistency checks for the obtained solutions and clarify the role played by the different parameters of the model.

The outline of this paper is as follows: In Section II, we present the two types of quintessence models that will be investigated, and we briefly review the formalism necessary to determine the Klein–Gordon equation in a Kiselev black hole background. In the same section, we decouple the angular and radial equations and solve the angular equation. In Section III, we solve the radial equation for both cases, obtain the quasi-energy spectrum, and analyze the Hawking radiation and Hawking temperature at the event horizon. In this section, we also examine, through graphical analysis, how dark energy influences the radiation observed in the region outside the event horizon. Finally, in Section IV, we present our conclusions.

\section{Scalar particle in the Kiselev-anti-de-Sitter spacetime}

Let the metric of the cosmological Kiselev black hole be given by \cite{gharderi}
\begin{eqnarray}\label{metricakiselevcosmo}
	ds^2 &=& \left(1-\frac{2M}{r} -\frac{\Lambda}{3} r^2- \frac{\alpha}{r^{3\omega_{0} + 1}}\right)dt^2 - \left(1-\frac{2M}{r}-\frac{\Lambda}{3} r^2 - \frac{\alpha}{r^{3\omega_{0} + 1}}\right)^{-1}dr^2 \nonumber\\\\ &&- r^2 d\theta^2 -r^2\sin^{2}\theta d\phi^2,\nonumber
\end{eqnarray}
where $M$ is the mass of the black hole, $\Lambda$ is the cosmological constant and $-1 < \omega_{0} < -1/3$, with $\alpha$ being a constant associated with the anisotropic fluid (quintessence-like). In this work, we will investigate the case for $\omega_{0} = -2/3$, thus the line element becomes
\begin{eqnarray}\label{metricageral}
	ds^2 &=& \left(1-\frac{2M}{r} -\frac{\Lambda}{3} r^2- \alpha r\right)dt^2 - \left(1-\frac{2M}{r}-\frac{\Lambda}{3} r^2 - \alpha r\right)^{-1}dr^2 \nonumber\\\\ &&- r^2 d\theta^2 -r^2\sin^{2}\theta d\phi^2.\nonumber
\end{eqnarray}

The horizons will be given by
\begin{equation}
	1 - \frac{2M}{r} -b r^2 -\alpha r  = 0 \hspace{0.5cm}\rightarrow \hspace{0.5cm}  -\frac{b(r-r_h)(r-r_1)(r-r_2)}{r} = 0,
\end{equation}
where $b = \Lambda/3$ and
\begin{eqnarray}
	\left. \begin{array}{l}
		\displaystyle r_h = -\frac{\alpha}{3b} + \frac{2}{3b}\sqrt{\alpha^2 + 3b}\cos\left[\frac{1}{3}\cos^{-1}\left(\frac{3q}{2p}\sqrt{\frac{-3}{p}}\right) - \frac{2\pi}{3}\right]\\\\
		\displaystyle r_1 = -\frac{\alpha}{3b} + \frac{2}{3b}\sqrt{\alpha^2 + 3b}\cos\left[\frac{1}{3}\cos^{-1}\left(\frac{3q}{2p}\sqrt{\frac{-3}{p}}\right)\right]\\\\
		\displaystyle r_2 = -\frac{\alpha}{3b} + \frac{2}{3b}\sqrt{\alpha^2 + 3b}\cos\left[\frac{1}{3}\cos^{-1}\left(\frac{3q}{2p}\sqrt{\frac{-3}{p}}\right) - \frac{4\pi}{3}\right]
	\end{array}\right.,
\end{eqnarray}
where $r_h + r_1 + r_2 = -\alpha/b$ and $r_1 > r_h > r_2$, with $p = -1/b - \alpha^2/(3b^2)$ and $q = (2\alpha^3/b^3 + 9\alpha/b^2 + 54M/b)/27$. For all three horizons to be real, the condition $4p^3 + 27q^2 < 0$ must be satisfied. Thus, $r_{h}$ represents the event horizon, $r_{1}$ the outer cosmological horizon, and $r_2$ the inner cosmological horizon. Furthermore, in the limit $\alpha \rightarrow 0$, we recover the values of $r_{h}$, $r_{1}$, and $r_2$ corresponding to the anti-de Sitter black hole \cite{dalpradesitter}. In the Kiselev black hole limit, with $\Lambda\rightarrow 0$ or $b\rightarrow 0$, we recover the event horizon $r_h = (1-\sqrt{1-8\alpha M})/2\alpha$ and $r_1 = (1+\sqrt{1-8\alpha M})/2\alpha$, which is commonly referred to as the quintessence-like horizon, while $r_2 \rightarrow -\infty$. Finally, in the limit $\alpha \rightarrow 0$ and $\Lambda\rightarrow 0$, we recover the event horizon of the Schwarzschild black hole, $r_{h} = 2M$.

To examine the dynamics of a relativistic scalar particle, we start from the Klein--Gordon equation, adopting natural units $\hbar = c = 1$. We consider a general spacetime described by the metric $ds^2 = g_{\mu\nu}dx^{\mu}dx^{\nu}$, for which the Klein--Gordon equation can be written as
\begin{equation}\label{kgmetricacurva}
\frac{1}{\sqrt{-g}}\partial_{\mu}(\sqrt{-g}g^{\mu\nu}\partial_{\nu}\Psi) = - m_{0}^{2}\Psi
\end{equation}
where $m_{0}$ is the mass of the scalar particle, with  $g = {\rm det}(g_{\mu\nu})$ where $g_{\mu \nu}$ and $g^{\mu \nu}$ are the metric tensor and its inverse, respectively, and we use the Einstein convention sum with repeated Greek indices run from 0 to 3. From (\ref{metricageral}) we identify the metric tensor $g_{\mu \nu}$,
\begin{equation}\label{tensormetrico}
g_{\mu\nu}= {\rm diag}\left[1-\frac{2M}{r} - b r^2 -\alpha r,-\left(1-\frac{2M}{r} - b r^2 -\alpha r\right)^{-1},-r^2,-r^2\sin^2\theta\right],
\end{equation}
thus the inverse metric tensor is,
\begin{equation}
g^{\mu\nu} = {\rm diag}\left[\left(1-\frac{2M}{r} - b r^2 -\alpha r\right)^{-1},-\left(1-\frac{2M}{r} - b r^2 -\alpha r\right),-\frac{1}{r^2},-\frac{1}{r^2\sin^2\theta}\right].
\end{equation}

Using $g = -r^4\sin^2(\theta)$ and (\ref{tensormetrico}), we write the Klein-Gordon equation in a Kiselev-anti-de-Sitter black hole background as, 
\begin{eqnarray}\label{kgequationfinal}
\left\{h^{2}(r)\left[\frac{\partial^2}{\partial r^2} + \left(\frac{2}{r} + \frac{h^{'}(r)}{h(r)}\right)\frac{\partial}{\partial r}\right] + \frac{h(r)}{r^2} \hat{L}^{2} - m_{0}^{2} h(r)- \frac{\partial^2}{\partial t^2}\right\}\Psi  = 0,
\end{eqnarray}
where $h(r) = 1 - 2M/r -br^2 - \alpha r$ with $h^{'}(r)$ is its the first derivative, and the square angular momentum operator $\hat{L}^{2}$ is given by
\begin{equation}\label{angularoperator}
\hat{L}^{2} = \frac{\partial^2}{\partial \theta^2} + \cot(\theta)\frac{\partial}{\partial \theta} + \frac{1}{\sin^2(\theta)} \frac{\partial^2}{\partial \phi^2},
\end{equation}
and its eigenfunctions are the well-known spherical harmonics $Y^{m}_{l}(\theta,\phi)$ \cite{arfken}, with $\hat{L}^{2} Y^{m}_{l}(\theta,\phi) = -l(l+1) Y^{m}_{l}(\theta,\phi)$, where $l$ is the orbital angular momentum and $m$ is the quantum number associated with the magnetic moment. Due to the spherical symmetry of the problem, we can write $\Psi(r,\theta,\phi,t) = R(r) Y^{m}_{l}(\theta,\phi) e^{-i \omega t}$, with $\omega$ being the energy, and by substituting this into (\ref{kgequationfinal}) we obtain the radial equation.
\begin{eqnarray}\label{radialequationgeral}
	\left\{h^{2}(r)\left[\frac{d^2}{d r^2} + \left(\frac{2}{r} + \frac{h^{'}(r)}{h(r)}\right)\frac{d}{d r}\right] - l(l+1) \frac{h(r)}{r^2}  - m_{0}^{2} h(r) + \omega^2\right\}R(r)  = 0.
\end{eqnarray}

Thus, by substituting the expression for $h(r)$, the equation above can be solved to determine the radial wave function $R(r)$, as will be shown in the next section.

\section{Radial wave equation exact solution}

In this section, we solve the radial equation and investigate some properties of the dynamics of a scalar particle in the background of a Kiselev--anti-de Sitter black hole. Using (\ref{radialequationgeral}), we obtain
\begin{eqnarray}\label{radialequation2}
	\left[\frac{d^2}{d r^2} + \left(\frac{1}{r} + \frac{1}{r-r_h} + \frac{1}{r-r_1} + \frac{1}{r-r_2}\right)\frac{d}{d r} + \frac{l(l+1)}{b r(r-r_{h})(r-r_{1})(r-r_{2})} + \nonumber \right. \\\\ \left.  \frac{m_{0}^{2}}{b}\frac{r}{ (r-r_{h})(r-r_{1})(r-r_{2})}  + \frac{\omega^2}{b^2}\frac{ r^2}{(r-r_{h})^2(r-r_{1})^2(r-r_{2})^2}\right]R(r) = 0.\nonumber
\end{eqnarray}

To solve the differential equation above, we perform the substitution $x = 1 - r/r_{h}$, thus we obtain
\begin{eqnarray}\label{radialequation2x}
	\left\{\frac{d^2}{d x^2} + \left(\frac{1}{x} + \frac{1}{x-1} + \frac{1}{x-a} + \frac{1}{x-a_1}\right)\frac{d}{d x} + \left[\frac{l(l+1)}{b r_{h}^2} + \frac{m_{0}^2}{b}(x-1)^2 -  \nonumber \right.\right. \\\nonumber\\ \left.\left.  \frac{2\omega^2(a a_1 - a_1 - a)}{b^2 r_{h}^2 a^3 a_{1}^3}(x-1)(x-a)(x-a_1) - \frac{2\omega^2(a-1)(a(a-2) +a_1)}{b^2 r_{h}^2 a^3 (a-a_{1})^3}x(x-1)(x-a_1) \right. \right.\nonumber \\\\ \left. \left. + \frac{2\omega^2 (a_1-1)(a_1(a_1-2) + 2)}{b^2r_{h}^2 a_{1}^3(a-a_1)^3}x(x-1)(x-a) \right]\frac{1}{x(x-1)(x-a)(x-a_1)} +  \left(\frac{\omega}{b r_h a a_1}\right)^2\frac{1}{x^2}   \right. \nonumber \\\nonumber \\ \left. + \left(\frac{\omega (a-1)}{b r_{h} a(a-a_1)}\right)^2\frac{1}{(x-a)^2}  + \left(\frac{\omega (a_1 -1)}{b r_{h} a_1 (a-a_1)}\right)^2\frac{1}{(x-a_1)^2} \right\}R(x)  = 0,\nonumber
\end{eqnarray}
where $a = 1-r_1/r_h$ and $a_1 = 1- r_2/r_h$. The solution of the differential equation in (\ref{radialequation2x}) can be obtained in terms of a generalized hypergeometric function via the ansatz $R(x) = x^{s_{1}}(x-a)^{s_{2}} (x-a_1)^{s_{3}} f(x)$ for eliminate the inverse quadratic terms $1/x^2$, $1/(x-a)^{2}$  and $1/(x-a_1)^{2}$ from the equation. Aplicando o ansatz em (\ref{radialequation2x}), we obtain
\begin{eqnarray}\label{radialequation2f}
	\left\{\frac{d^2}{d x^2} + \left(\frac{2s_{1}+1}{x} + \frac{1}{x-1} + \frac{2s_{2}+1}{x-a} + \frac{2s_{3}-2}{x-a_1}\right)\frac{d}{d x} + \nonumber\right. \\\\ \left. \frac{\mu_1 x^2 + \mu_2 x + \mu_3}{x(x-1)(x-a)(x-a_1)}\right\}f(x)  = 0,\nonumber
\end{eqnarray}
where 
\begin{eqnarray*}
 \mu_{1} &=&  3(s_{1} + s_{2} + s_{3}) + 2(s_1 s_2 + s_1 s_3 + s_2 s_3)  + \frac{m_{0}^{2}}{b} + \frac{2\omega^2(a a_1-a_1 - a)}{b^2 r_{h}^2 a^3 a_{1}^3}(a + a_1 + 1) \\\\ &&
  + \frac{2\omega^2(a-1)(a_1+1)[a(a-2)+a_1]}{b^2 r_{h}^2 a^3 (a-a_{1})^3}(a + a_1 + 1) - \frac{2\omega^2(a_1-1)(a+1)[a_1(a_1-2)+a]}{b^2 r_{h}^2 a_{1}^3 (a-a_{1})^3}\times \\\\ && (a + a_1 + 1),
\end{eqnarray*}
\begin{eqnarray*}
 \mu_{2} &=&  -2s_1(a_1 + a + a_1) - 2s_2(a_1 + 1) -2s_3(a+1) - \frac{2m_{0}^{2}}{b} - \frac{2\omega^2(a a_1-a_1 - a)}{b^2 r_{h}^2 a^3 a_{1}^3}\times \\\\ && (a a_1 + a_1 + 1) - \frac{2\omega^2 a_1(a-1)[a(a-2)+a_1]}{b^2 r_{h}^2 a^3 (a-a_{1})^3}(a + a_1 + 1) + \frac{2\omega^2 a(a_1-1)[a_1(a_1-2)+a]}{b^2 r_{h}^2 a_{1}^3 (a-a_{1})^3} \\\\ &&
  -2s_1 s_2 (a_1 + 1) - 2s_1 s_3 (a+1) - 2s_2 s_3 ,
\end{eqnarray*}
\begin{eqnarray*}
 \mu_{3} = s_1(a_1 + a+ a a_1) + a_1 s_2 + a s_3 + \frac{l(l+1)}{b r_{h}^2} + \frac{m_{0}^2}{b} + \frac{2\omega^2 (a a_1 -a-a_1)}{b^2 r_{h}^2 a^2 a_{1}^2},
\end{eqnarray*}
with $s_{1} =  \pm i\omega/(br_{h} a a_1)$, $s_2 = \pm i\omega(a-1)/( b r_{h}a(a-a_1)) $ and $s_{3} = \pm i\omega (a_1 -1)/(b r_{h}a_1(a-a_1))$. The equation (\ref{radialequation2f}) is analogous to the generalized hypergeometric equation, which has four finite singularities at $z = (0,1,a,a_1)$ and one at infinity $z \rightarrow \infty$, and in a generalized form is given by \cite{ishkhangen}
\begin{equation}\label{heunequation}
	y''(z) + \left(\frac{\gamma}{z} + \frac{\delta }{z-1} + \frac{\epsilon}{z-a} + \frac{\epsilon_1}{z-a_1}\right)y'(z) + \frac{\eta\beta z^2 - \theta_1 z - \theta_0}{z(z-1)(z-a)(z-a_1)}y(z) = 0,
\end{equation}
where $y_{01}$ and $y_{02}$ are constants and the regularity of the singularity at infinity is ensured by the Fuchsian condition $1+\eta + \beta = \gamma + \delta + \epsilon + \epsilon_1$. The function $y(z)$ is given by 
\begin{equation}
	y(z) = y_{01} u(a,a_1,\gamma,\eta,\beta,\theta_1,\theta_0;z) + y_{02} z^{1-\gamma}u(a,a_1,\gamma',\eta',\beta',\theta_{1}',\theta_{0}';z),
\end{equation}
with $\gamma' = 2-\gamma$, $\eta'\beta' = (\delta + \epsilon + \epsilon_1)(1-\gamma) + \eta\beta$, $\theta_{1}' = [\delta(a+a_1) +\epsilon(1+a_1) + \epsilon_1 (1+a)](1-\gamma) + \theta_1$ and $\theta_{0}' = [\delta a a_1 + \epsilon a_1 + \epsilon_1 a](1-\gamma) + \theta_0$. The generalized hypergeometric function is defined by 
\begin{equation}
	u(a,a_1,\gamma,\eta,\beta,\theta_1,\theta_0;z) = \sum_{k=0}^{\infty} \tau_{k}z^{k},
\end{equation}
where the coefficients $\tau_{k}$ are obtained through the four-term recurrence relation given by
\begin{equation}\label{recorrence}
	A_{k+1}\tau_{k+1} + B_{k}\tau_{k} + C_{k-1}\tau_{k-1} + D_{k-2}\tau_{k-2} = 0,
\end{equation}
fpr $k \geq 0$, with $\tau_{-2} = \tau_{-1} = 0$, $\tau_{0} = 1$, and
\begin{eqnarray}
	\left. \begin{array}{l}
		\displaystyle	A_{k} = a k(\gamma + k -1) \\
		\displaystyle	B_{k} = \frac{\theta_0}{a_1} - \left[(\gamma+ k -1)\left(\frac{a}{a_1} + a+1\right) + \frac{a}{a_1}\epsilon_1 + \epsilon + a \delta\right]k\\\\
		\displaystyle C_{k} = \frac{\theta_1}{a_1} +  \left[(\eta + \beta + k )\left(\frac{a+1}{a_1} +1\right) -\frac{(\delta + a\epsilon)}{a_1} - \epsilon_1 \right]k \\\\\
		\displaystyle D_{k} = -\frac{1}{a_1}(k+\eta)(k + \beta)
	\end{array}\right..
\end{eqnarray}

Applying (\ref{recorrence}) for $k=0,1,2,\ldots,n$, we obtain a system of $n+1$ homogeneous equations, which can be written as
\begin{eqnarray}\label{relacaomatricial}
	\left(\begin{array}{cccccc}
		B_{0} & A_1 & 0 & 0 & 0 & \ldots \\
		C_0 & B_1 & A_2 & 0 & 0 & \ldots \\
		D_{0} & C_1 & B_2 & A_3 & 0 & \ldots\\
		0 &  D_{1} & C_2 & B_3 & A_4  & \ldots\\
		\vdots & \vdots & \vdots & \vdots & \vdots
	\end{array}\right) \left(\begin{array}{c}
		\tau_0\\
		\tau_1\\
		\tau_2\\
		\tau_3\\
		\vdots
	\end{array}
	\right) = \left(\begin{array}{c}
		0\\
		0\\
		0\\
		0\\
		\vdots
	\end{array}
	\right).
\end{eqnarray}

In our case, we identify $\eta \beta  = \mu_1$, $\theta_1 = -\mu_2$, $\theta_0 = -\mu_3$, $\gamma = 2s_1 + 1$, $\delta =  1$, $\epsilon = 2s_2 + 1$ and $\epsilon_1 = 2s_3+1$. Thus, the radial wave function $R(x)$ will be given by 
\begin{eqnarray}\label{radialwfgeral2}
	R(x) &=& x^{s_{1}}(x-a)^{s_{2}} (x-a_1)^{s_{3}}[ y_{01} u(a,a_1,\gamma,\eta,\beta,\theta_1,\theta_0;x) +\nonumber \\&& y_{02} x^{-2s_1}u(a,a_1,\gamma',\eta',\beta',\theta_{1}',\theta_{0}';x),
\end{eqnarray}
where $-\infty < x \leq 0 $.

\subsection{Quasibound states and quasispectrum}

In this section, we derive the energy quasispectrum. It should be noted that genuine bound states cannot be defined over the complete range of $x$, because the radial wave function in (\ref{radialwfgeral2}), although regular, does not vanish at either $x=0$ or in the limit $x\rightarrow-\infty$. We therefore impose the weaker requirement that $R(x)$ remain finite at both endpoints. Under these conditions, the resulting solutions describe quasibound states characterized by complex resonant frequencies, which collectively form the quasispectrum. The required behavior at $x\rightarrow-\infty$ is ensured by imposing a termination condition on the power-series expansion, reducing it to a polynomial. Accordingly, the infinite series defining the function $\eta$ must terminate so that the radial solution remains finite and does not diverge as $x\rightarrow-\infty$.
 Therefore, by truncating the series at a given $k = n+2$, with $n \geq 0$ of the first function $u$ in (\ref{radialwfgeral2}) and doing $y_{02} = 0$, we obtain the following conditions
\begin{equation}
	i)\;\;\; D_{n} = 0 \hspace{1cm} {\rm and} \hspace{1cm}
	ii) \;\;\; \Delta_{n+3} = \left|\begin{array}{ccccc}
		B_{0} & A_{1} & 0 & 0 & \ldots\\
		C_{0} & B_{1} & A_{2} & 0 &\ldots\\
		D_{0} & C_{1} & B_{2} & A_{3} & \ldots\\
		\vdots & \vdots & \vdots  & \vdots  & \vdots\\
		0 & 0 & 0 & D_{n-2} & C_{n-1}
	\end{array}\right| = 0.
\end{equation}

By applying the first condition $D_{n} = 0$, we have $\beta =  -n$ or $\eta = -n$, and using $\eta \beta = \mu_1$ and the condition $1+\eta + \beta = \gamma + \delta + \epsilon + \epsilon_1$,  we obtain
\begin{eqnarray}
	n^2 + (2s_1 + 2s_2 + 2s3 + 3)n + \mu_1 = 0,
\end{eqnarray}
and using the value of the constants above, the quasispectrum $\omega$ will be given by
\begin{equation}\label{quasispectrum}
	\omega_n = -\frac{ibr_h a a_1}{4}(2n+3) \pm \frac{b r_h a a_1}{4}\sqrt{\frac{4m_{0}^2}{b}-9 }, 
\end{equation}
for $n \geq 0$, where we use $s_{1} = -i \omega /(b r_h a a_1)$, $s_{2} = -i\omega (a-1)/(b r_h a (a-a_1) )$, and $s_3 = i\omega (a_1 - 1)/(b r_h a_1 (a-a_1)$) to ensure consistency with the results obtained in the asymptotic cases, as will be discussed below. Thus, we observe that, depending on the values of $m_{0}$, $b$, and $n$, the energy quasispectrum may have both real and imaginary components or may be purely imaginary. The real component is not quantized and exists only when the condition $m_{0}^2 \geq 9b/4$ is satisfied, with $b = \Lambda/3$. In the figures below, we plot the energy quasispectrum for the ground state to investigate how $\alpha$ and $\Lambda$ affect the particle dynamics. In both figures, we set $m_{0} = 1$.

\graphicspath{{figuras/}}

\begin{figure}[!htb]
	\centering
	\includegraphics[scale={0.7}]{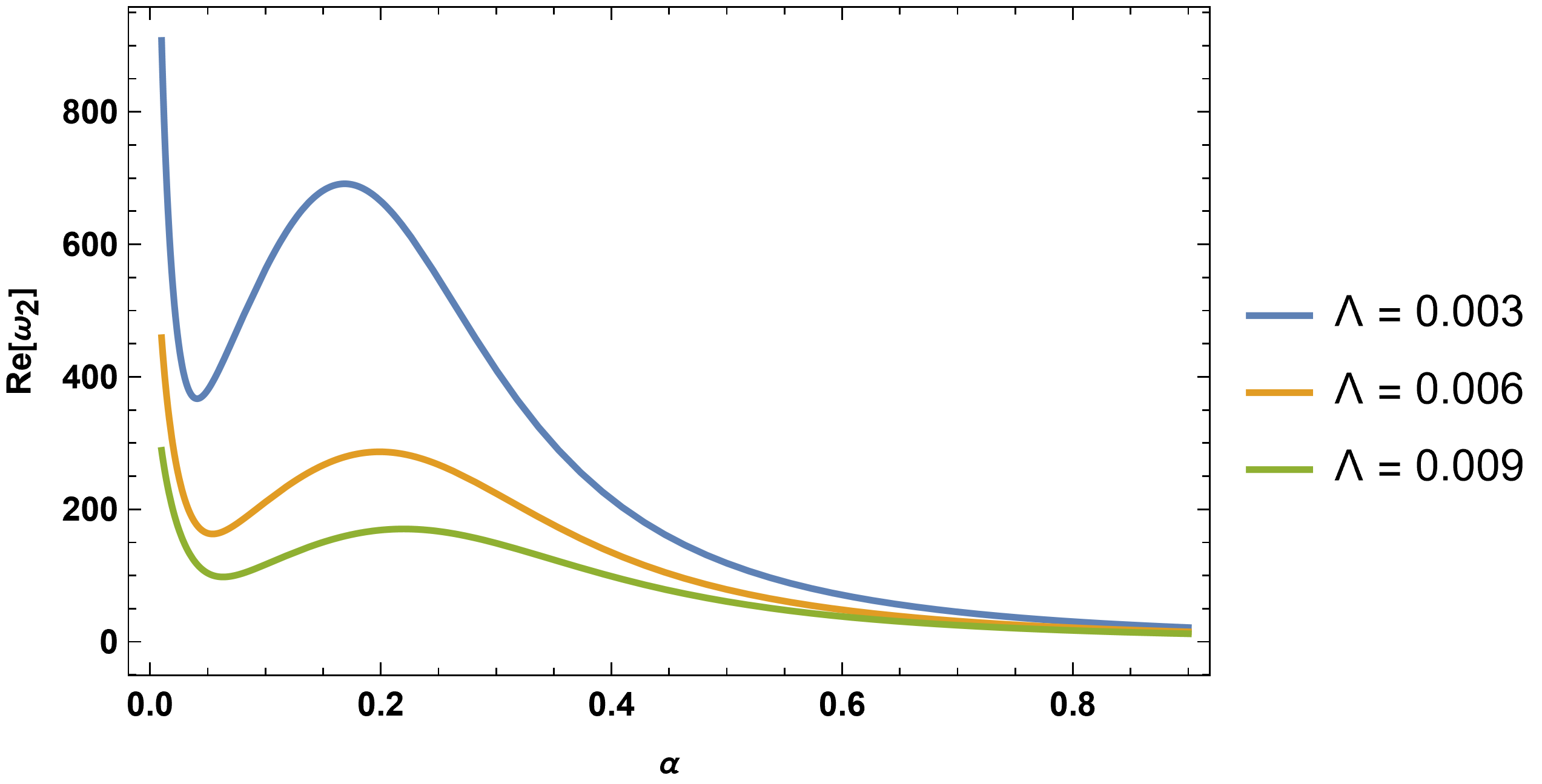}
	\caption{Representation of the real component of the quasi-energy spectrum for the ground state $(n = 0)$ as a function of $\alpha$ for different values of $\Lambda$.}
	\label{figuraenergia1}
\end{figure}

\graphicspath{{figuras/}}

\begin{figure}[!htb]
	\centering
	\includegraphics[scale={0.7}]{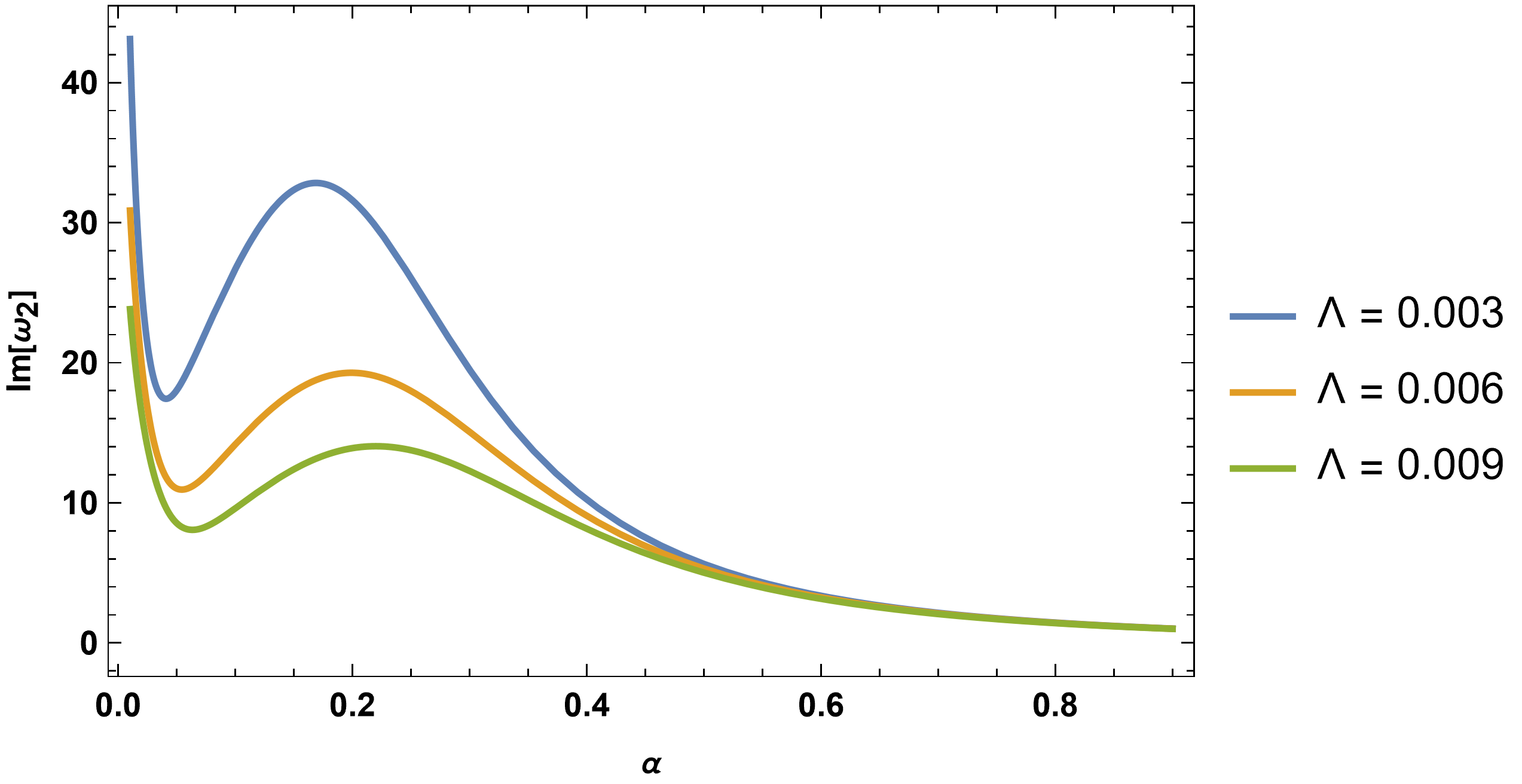}
	\caption{Representation of the imaginary component of the quasi-energy spectrum for the ground state $(n = 0)$ as a function of $\alpha$ for different values of $\Lambda$.}
	\label{figuraenergia2}
\end{figure}

As shown in Figs. (\ref{figuraenergia1}) and (\ref{figuraenergia2}), decreasing the values of $\alpha$ and $\Lambda$ leads to higher-energy particle states. This behavior is observed for both the real and imaginary components of the energy quasispectrum. We also note that the real component is approximately one order of magnitude larger than the imaginary component.

Finally, using the second condition given by $\Delta_{n+3} = 0$, we can obtain the value of a physical parameter such as, for example, the angular momentum $l$ for each specific value of $n$. In other words, these values will be quantized and will assume complex values due to parameters with complex values such as $s_{1}$, $s_{2}$ and $s_3$. For the ground state, for example, we have $n = 0$; therefore, by imposing $\Delta_{3} = 0$, we obtain
\begin{equation}
	\left|\begin{array}{ccc}
		B_{0} & A_{1} & 0 \\
		C_{0} & B_{1} & A_{2} \\
		0 & C_{1} & B_{2}
	\end{array}\right| = B_{0}B_{1}B_{2} + C_{1}A_{2}B_{0} - B_{2}C_{0}A_{1}  = 0,
\end{equation}
Thus, by specifying the parameter values, the equation can be solved numerically, allowing us, for instance, to determine the complex values of the angular momentum.

The radial wave function will be 
\begin{eqnarray}\label{Rfinal}
	R_n(x) &=&  x^{s_{1}}(x-a)^{s_{2}} (x-a_1)^{s_{3}}\sum_{k = 0}^{n} \tau_{k}x^{k}.
\end{eqnarray}

Finally, we must verify that the wave function remains finite as $x\rightarrow-\infty$. To this end, we consider the generalized hypergeometric-type equation (\ref{heunequation}) in the vicinity of the singular point at infinity, where two linearly independent solutions exist. Following the procedure commonly adopted for Heun-type equations \cite{ronveaux}, we expand the function $u$ in this region, which leads to the following asymptotic series
\begin{equation}\label{Heunassintotico}
	u(a,a_1,\gamma,\eta,\beta,\theta_1,\theta_0;x)] \approx c_{1}x^{-\rho_+} + c_{2}x^{-\rho_{-}}
\end{equation}
where we retain only the leading terms of the asymptotic power series, with $\rho_{\pm}=s_1+s_2+ + s_3 + 5/2\pm i\chi$ and $\chi = \sqrt{\mu_1 + \left(s_1+s_2 + s_3  +5/2\right)^2}$. Thus, from equations (\ref{radialwfgeral2}) and (\ref{Heunassintotico}), the radial wave function far from the Kiselev-anti-de-Sitter black hole, i.e., in the limit $|x|\rightarrow\infty$, is given by
\begin{equation}\label{radialfunction1}
	R(x) \approx \frac{1}{x^{5/2}}[c_{1}x^{-i\chi} + c_{2}x^{i\chi}] =  \frac{1}{x^{5/2}}[c_{1} e^{-i\chi \ln(x)} + c_{2} e^{i\chi \ln(x)}].
\end{equation}

Thus, the radial wave function can be written as
\begin{equation}
	R(x) \approx c_{\lambda} \frac{1}{x^{5/2}}\sin[\chi {\rm ln}(x) + \sigma_{\lambda}(\omega)],
\end{equation}
where $\sigma_{\lambda}(\omega)$ is the phase shift. We then verify that the wave function remains finite in the limit $x\rightarrow-\infty$, as required. The parameter $\sigma_{\lambda}$ can be determined analogously by following the procedure developed by Abramov \textit{et al.} \cite{abramov} for the Heun function.

\subsection{Hawking's radiation}

In this section, we calculate the Hawking radiation through the rate of particles that escape from the black hole across the event horizon. To do this, we first determine the wave function in the region of the event horizon with $r \approx r_{h}$ or $x \approx 0$. Thus, from (\ref{radialwfgeral2}) we have
\begin{equation}
	R(r\approx r_{h}) \approx a_{1} (r-r_{h})^{s_{1}} + a_{2}(r-r_{h})^{-s_{1}},
\end{equation}
where $a_1$ and $a_2$ absorb all constant factors, and we use
$u(a,a_1,\gamma,\eta,\beta,\theta_1,\theta_0;0)=1$, and the time-dependent wave function for $r \approx r_{h}$ will be
\begin{equation}
	\psi(r,t) \approx [a_{1}(r-r_{h})^{s_{1}} + a_{2}(r-r_{h})^{-s_{1}}]e^{-i\omega t},
\end{equation}
where $s_{1} = -i\omega/(b r_h a a_1)$, with $\text{Im}[s_1] > 0$, because $a<0$. We define $\psi_{in}$ and $\psi_{out}$ as the wave functions describing the modes entering and leaving the event horizon, respectively, with
\begin{eqnarray}
	\left\{\begin{array}{l}
		\psi_{in} = (r-r_{h})^{-s_1}e^{-i\omega t}\\
		\psi_{out} = (r-r_{h})^{s_1}e^{-i\omega t}\\
	\end{array}
	\right. ,
\end{eqnarray} 

Now, we perform the transformation to the Eddington–Finkelstein coordinate, which is given by
\begin{equation}
	dr' = -\frac{r}{b(r-r_{h})(r-r_{1})(r-r_2)}dr \approx -\frac{dr}{b r_{h}a a_1(r-r_{h})},
\end{equation}
where we have applied the condition $r \approx r_{h}$. Thus, by integrating both sides we obtain
\begin{equation}
	r' = -\frac{1}{b r_{h}a a_1}\ln(r-r_{h}) \hspace{0.7cm}\rightarrow \hspace{0.7cm} r-r_{h} = e^{-b r_{h}a a_1 r'},
\end{equation}
and introducing now the variable $v = t + r'$, we finally obtain
\begin{eqnarray}
	\left\{\begin{array}{l}
		\psi_{in} = e^{-i\omega v}\\
		\psi_{out} = e^{-i\omega v} (r-r_{h})^{is'}
	\end{array}
	\right. ,
\end{eqnarray} 
where $s' = -2\omega/(b r_{h}a a_1)$ and since we want to calculate the rate of particles escaping from the event horizon, we will use only the wave function $\psi_{out}$ in the regions $r>r_{h}$ and $r<r_{h}$. For the region $r<r_{h}$, we need to perform an analytic continuation due to the pole at $r = r_{h}$. Thus, by considering a contour in the lower complex semi-plane, we obtain $r-r_{h} = (r_{h}-r)e^{-i\pi}$, and therefore we have that
\begin{eqnarray}
	\left\{\begin{array}{l}
		\psi_{out}(r<r_{h}) = e^{-i\omega v}(r_{h}-r)^{is'}e^{\pi s'} \\
		\psi_{out}(r>r_{h}) = e^{-i\omega v} (r-r_{h})^{is'}\\
	\end{array}
	\right. ,
\end{eqnarray} 
and using the formulation of Sannan \cite{sannan}, the decay rate will be given by
\begin{equation}
	\Gamma = \left|\frac{\psi_{out}(r>r_{h})}{\psi_{out}(r<r_{h})}\right|^2 = \exp\left(\frac{4\pi \omega}{b r_{h} a a_1}  \right),
\end{equation}
where $a = (r_{h}-r_{1})/r_{h} < 0$. With that, the Hawking radiation will be
\begin{equation}
	N_{H} = \frac{\Gamma}{1-\Gamma} = \left[\exp\left( -\frac{4\pi \omega}{b r_{h} a a_1} \right)-1\right]^{-1}.
\end{equation}

By comparing this result with the Bose–Einstein statistics, which is given by $N_{BE} = [\exp(\omega/k_{B}T)-1]^{-1}$, we obtain that the Hawking temperature is given by $\omega/(k_{B}T_{H}) = -4\pi \omega/(b r_{h} a a_1) $, and solving it, we obtain that the Hawking temperature is
\begin{equation}\label{temperatura}
	T_{H} = -\frac{b r_{h} a a_1}{4\pi k_{B}} = \frac{b(r_1 - r_h)(r_h - r_2)}{4\pi k_{B}r_{h}},
\end{equation}
we observe that the Hawking temperature no depends on the energy of the particle that escapes from the event horizon. We also note that the Hawking temperature obtained here coincides with that derived using other methods, such as the surface gravity approach, for which the Hawking temperature is defined as $T_{H} = f'(r_{h})/(4\pi k_{B}) = b(r_{1}-r_{h})(r_h - 2_2)/(4\pi k_{B} r_{h})$, where $f(r) = 1-\alpha r - 2M/r - br^2$. We therefore observe that the anisotropic fluid, cosmological constant  and the particle's energy directly affect the Hawking radiation, whereas the black hole temperature is influenced solely by the anisotropic fluid and cosmological constant. In the figure below, we analyze the behavior of the radiation for different values of $\Lambda$ and $\alpha$.

\graphicspath{{figuras/}}

\begin{figure}[!htb]
	\centering
	\includegraphics[scale={0.7}]{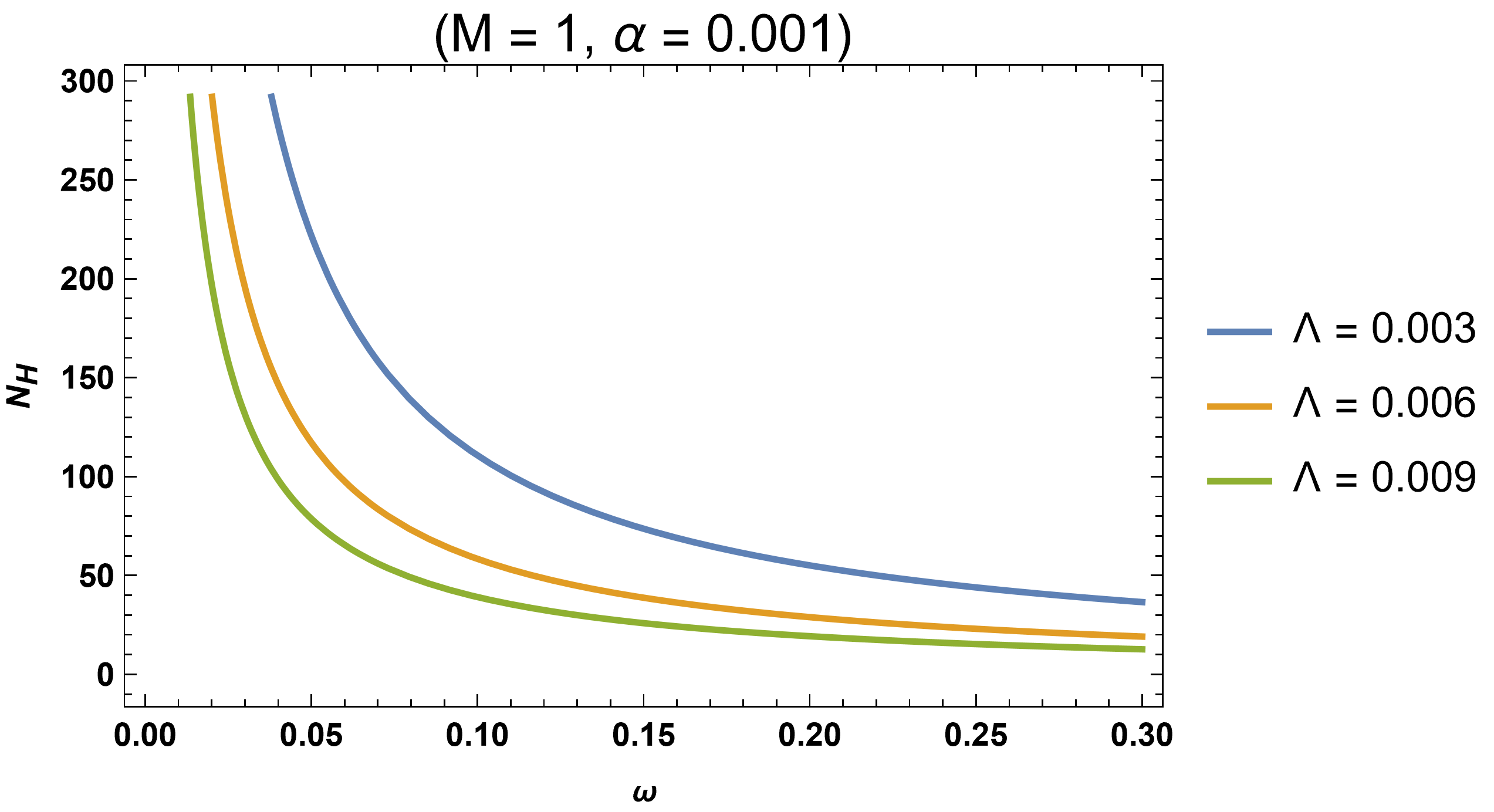}
	\caption{Representation of the behavior of Hawking radiation in terms of energy $\omega$ for different values of $\alpha$.}
	\label{figuraradiation1}
\end{figure}

\graphicspath{{figuras/}}

\begin{figure}[!htb]
	\centering
	\includegraphics[scale={0.7}]{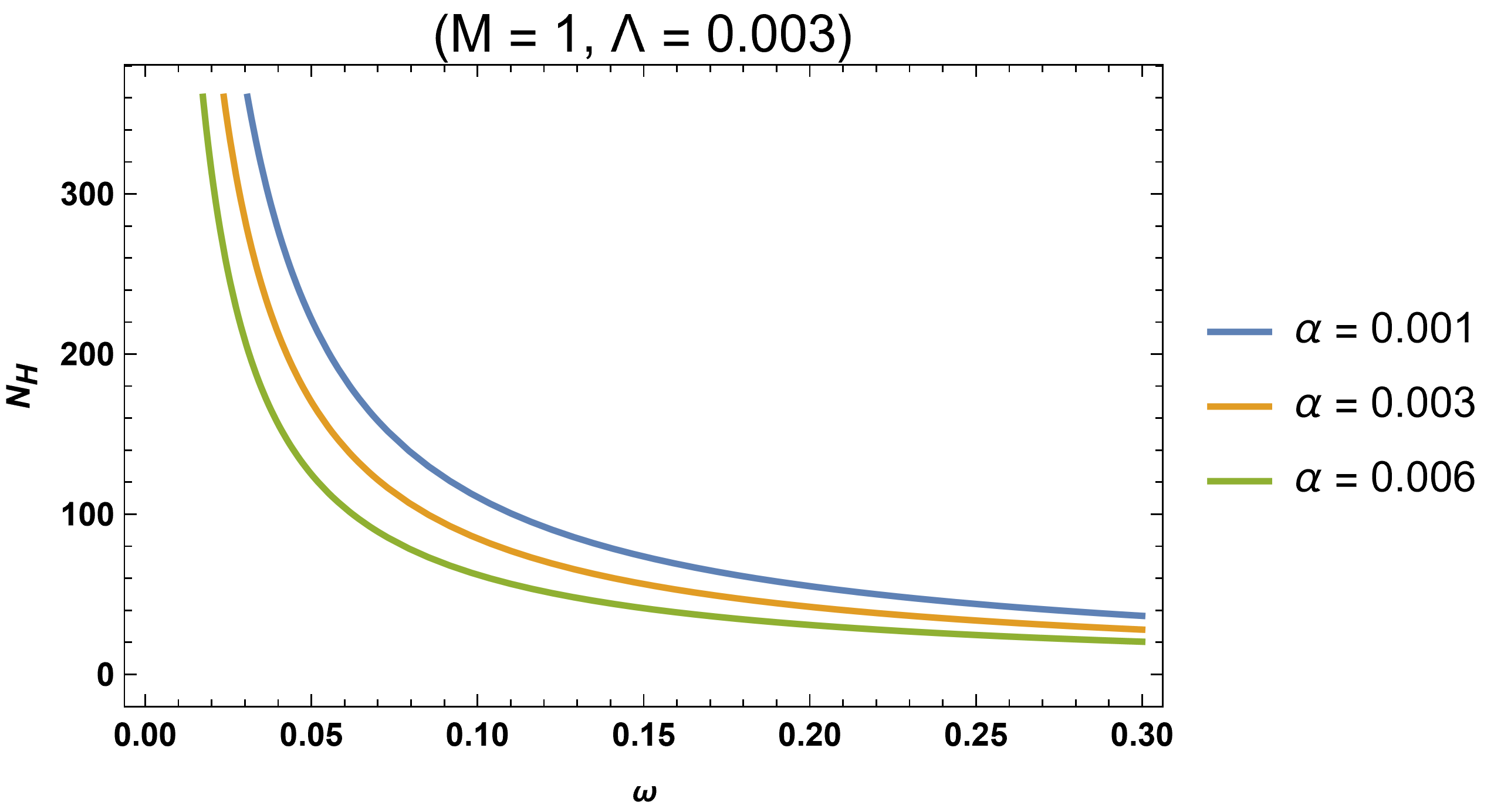}
	\caption{Representation of the behavior of Hawking radiation  in terms of energy $\omega$ for different values of $\alpha$.}
	\label{figuraradiation2}
\end{figure}

As illustrated in Figs. (\ref{figuraradiation1}) and (\ref{figuraradiation2}), lower values of $\Lambda$ and $\alpha$ are associated with a greater amount of radiation in the region outside the event horizon. Since black holes are generally expected to strongly suppress the escape of matter from their interior, this behavior may provide an indication of how dark energy-like effects, represented here by the cosmological constant and/or an anisotropic fluid, can influence the radiation emitted by the system. In this sense, our results suggest that increasing the contribution of the anisotropic fluid and/or the cosmological constant leads to a reduction in the amount of matter escaping from the black hole.

\subsection{Asymptotic cases}

\subsubsection{Kiselev back hole}

In this section, we examine several asymptotic limits. We first consider the Kiselev black hole limit, obtained by taking $b\rightarrow 0$ or $\Lambda \rightarrow 0$. In this regime, we must set $m_{0}=0$ to avoid a divergence in (\ref{quasispectrum}), corresponding to the massless scalar field case. Moreover, under this limiting procedure, we have $D_{k-2}\rightarrow 0$ in (\ref{recorrence}). Consequently, the differential equation in (\ref{radialequation2f}) reduces to a local Heun equation \cite{ronveaux}, while the recurrence relation in (\ref{recorrence}) takes the form
\begin{equation}
	A_{k+1}\tau_{k+1} + B_{k}\tau_{k} + C_{k-1}\tau_{k-1} = 0,
\end{equation}
for $k \geq 0$ and
\begin{eqnarray}
	\left. \begin{array}{l}
		\displaystyle	A_{k} = a k(\gamma + k -1) \\\\
		\displaystyle	B_{k} = \frac{\theta_0}{a_1} - \left[(\gamma+ k -1)\left( a+1\right) + \epsilon + a \delta\right]k\\\\
		\displaystyle C_{k} = \frac{\theta_1}{a_1} +  \left[\eta + \beta + k   - \epsilon_1 \right]k
	\end{array}\right..
\end{eqnarray}
with $\eta + \beta = \gamma + \delta + \epsilon + \epsilon_1 - 1$.

We now note that, in order for the power series to terminate at $k=n+1$, with $n\geq0$, it is necessary to impose the conditions $C_{n}=0$ and $\tau_{n+2}=\tau_{n+1}=0$. Therefore, we obtain $C_n = -\mu_2/a_1 + (2s_1 + 2s_2 +n + 2)n = 0.$ Solving this relation, we recover the corresponding value of the energy quasispectrum, which is given by
\begin{equation}\label{quasispequint}
	\omega_n = -\frac{i \alpha a (a-2)}{4(a-1)}(n+1) \pm \frac{i \alpha a (a-2)}{4(a-1)} \sqrt{4 + a[a(n+1)^2 - 4]},
\end{equation}
and in the limit $b\rightarrow 0$, we have that temos que $a_1 \rightarrow \infty$ and $b a_1 \rightarrow \alpha/r_{h}$, with
\begin{eqnarray}
	\left. \begin{array}{l}
		\displaystyle	r_{h} \rightarrow \frac{1 - \sqrt{1-8\alpha M}}{2\alpha}\\\\
		\displaystyle	r_{1} \rightarrow \frac{1 + \sqrt{1-8\alpha M}}{2\alpha}\\\\
		\displaystyle	r_2 \rightarrow -\infty
	\end{array}\right. .
\end{eqnarray}

Furthermore, the Hawking temperature in (\ref{temperatura}) reduces to
\begin{equation}
	T_{H} = \frac{\alpha (r_1 - r_h)}{4\pi k_{B} r_{h}}
\end{equation}
where $b(r_h - r_2) \rightarrow \alpha$. In this limit, we recover the results previously obtained for the Kiselev black hole \cite{dalpraquint1}.

\subsubsection{Anti-de-Sitter black hole }

The second limiting case we consider corresponds to the de Sitter black hole limit, obtained by taking $\alpha\rightarrow 0$. In this case, the differential equation retains the same form as that given in (\ref{radialequation2x}). Consequently, the energy quasispectrum remains unchanged and is still given by (\ref{quasispectrum}), while the Hawking temperature is likewise described by (\ref{temperatura}). The main modification arises in the locations of the horizons, which are now determined by
\begin{eqnarray}
	\left. \begin{array}{l}
		\displaystyle r_h =   \frac{2}{\sqrt{3b}}\cos\left[\frac{1}{3}\cos^{-1}\left(-M\sqrt{27b}\right) - \frac{2\pi}{3}\right]\\\\
		\displaystyle r_1 = \frac{2}{\sqrt{3b}}\cos\left[\frac{1}{3}\cos^{-1}\left(-M\sqrt{27b}\right)\right]\\\\
		\displaystyle r_2 = \frac{2}{\sqrt{3b}}\cos\left[\frac{1}{3}\cos^{-1}\left(-M\sqrt{27b}\right) - \frac{4\pi}{3}\right]
	\end{array}\right.,
\end{eqnarray}
where $r_h + r_1 + r_2 = 0$ and $r_1 > r_h > r_2$, with $p = -1/b $ and $q = 2M/b$. Thus, we recover the results previously obtained for the anti-de Sitter black hole \cite{dalpradesitter}.

\subsubsection{Schwarzschild black hole}

In the final case, we consider the Schwarzschild black hole limit by taking $\Lambda \rightarrow 0$ and $\alpha \rightarrow 0$. In this limit, the energy quasispectrum in (\ref{quasispequint}) reduces, as $\alpha \rightarrow 0$, to
\begin{equation}
	\omega_n = i\frac{(n+1)}{4M},
\end{equation}
where $a \rightarrow \infty$, $a\alpha \rightarrow -1/2M$, $r_{h} \rightarrow 2M$ and $r_{1} \rightarrow \infty$. Moreover, the Hawking temperature is given by

\begin{equation}
	T_{H} = \frac{1}{8\pi k_{B} M}.
\end{equation}

Thus, we recover the corresponding results for the Schwarzschild black hole \cite{bezerra1}.

\section{Conclusion}

In this work, we have investigated the dynamics of a scalar particle in a black hole spacetime surrounded by a cosmological constant and an anisotropic fluid. By solving the corresponding Klein--Gordon equation, we obtained the radial equation and analyzed its solutions in terms of a power-series expansion. By imposing the appropriate truncation conditions on the recurrence relation, we obtained the quasispectrum of energy and identified the conditions under which its real and imaginary components are present. In particular, depending on the values of the parameters $m_0$, $b$, and $n$, the quasispectrum can contain both real and imaginary contributions or can be purely imaginary. The real component is not quantized and exists only when $m_0^2\geq 9b/4$, with $b=\Lambda/3$. The numerical analysis in Figs. (\ref{figuraenergia1}) and (\ref{figuraenergia2}) showed that decreasing $\alpha$ and $\Lambda$ increases both components of the energy quasispectrum. We also found that the real component is approximately one order of magnitude larger than the imaginary component for the parameter values considered.

We have also investigated the radiation associated with the scalar field in the exterior region of the event horizon. The results indicate that smaller values of $\alpha$ and $\Lambda$ are associated with a larger amount of radiation outside the horizon, as we nota in Figs. (\ref{figuraradiation1}) and (\ref{figuraradiation2}). Conversely, increasing the contribution of the anisotropic fluid and/or the cosmological constant suppresses the radiation in the exterior region. This behavior illustrates how the surrounding matter content and the cosmological contribution can affect the dynamics of scalar fields in black hole spacetimes. 

As a consistency check, several relevant asymptotic limits were considered. First, by taking $b\rightarrow0$ or $\Lambda\rightarrow0$, we recovered the Kiselev black hole case. In this limit, the requirement $m_0=0$ leads to the massless scalar-field configuration, while the differential equation reduces to a local Heun equation. The corresponding truncation conditions reproduce the previously obtained energy quasispectrum and, together with the limiting behavior of the horizon parameters, recover the results for the Kiselev black hole \cite{dalpraquint1}. The Hawking temperature also reduces consistently to the corresponding expression in this limit.

We then considered the de Sitter black hole limit by setting $\alpha\rightarrow0$. The radial differential equation retains the same form, and consequently the expressions for the energy quasispectrum and Hawking temperature remain unchanged. The modification is instead reflected in the locations of the horizons. In this limit, the results previously obtained for the anti-de Sitter black hole are recovered \cite{dalpradesitter}. Finally, by simultaneously taking $\Lambda\rightarrow0$ and $\alpha\rightarrow0$, we obtained the Schwarzschild black hole limit. The corresponding expressions for the energy quasispectrum and Hawking temperature consistently reduce to the Schwarzschild results \cite{bezerra1}.

Overall, the analysis presented here shows that the quasispectrum and radiation of scalar particles are sensitive to both the cosmological contribution and the surrounding anisotropic matter. The consistency of the limiting cases with previously established black hole solutions provides an important check of the results. These findings may be useful for further investigations of scalar-field dynamics, quasibound states, and quantum effects in black hole spacetimes surrounded by nontrivial cosmological and matter configurations.

\section*{Acknowledgments}

AGMS gratefully acknowledges CNPq (grant number 309052/2023-8) for partial financial support. This study was funded by FAPERJ - Fundação Carlos Chagas Filho de Amparo à Pesquisa do Estado do Rio de Janeiro, Processess SEI 26/200.337/2024 and SEI 260003/021954/2025.

\section*{Data Availability Statement}

No Data associated in the manuscript.

\end{document}